\documentclass[manuscript=article, journal = nalefd]{achemso}
\usepackage{amsmath,amssymb}
\usepackage{hyperref}
\usepackage[per-mode=symbol]{siunitx}
\SectionNumbersOn 
\usepackage{pdfpages}

\author{Irene Sánchez Arribas}
\affiliation{%
Department of Electrical Engineering, School of Computation, Information and Technology, Technical University of Munich, 85748 Garching, Germany
}%

\author{Takashi Taniguchi}
\affiliation{International Center for Materials	Nanoarchitectonics, National Institute for Materials Science, Tsukuba, Ibaraki 305-0044, Japan}
\author{Kenji Watanabe}
\affiliation{Research Center for Functional Materials,
	National Institute for Materials Science, Tsukuba, Ibaraki 305-0044, Japan}

\author{Eva M. Weig}
\email{*eva.weig@tum.de}
\affiliation{%
	Department of Electrical Engineering, School of Computation, Information and Technology, Technical University of Munich, 85748 Garching, Germany
}%
\alsoaffiliation{%
	Munich Center for Quantum Science and Technology (MCQST), 80799 Munich, Germany
}%
\alsoaffiliation{%
	TUM Center for Quantum Engineering (ZQE), 85748 Garching, Germany
}%

\title
{Radiation pressure backaction on a hexagonal boron nitride nanomechanical resonator}

\begin{document}
	\newpage

	\begin{abstract}
		Hexagonal boron nitride (hBN) is a van der Waals material with excellent mechanical properties hosting quantum emitters and optically active spin defects, several of them being sensitive to strain. Establishing optomechanical control of hBN will enable hybrid quantum devices that combine the spin degree of freedom with the cavity optomechanical toolbox. In this letter, we report the first observation of radiation pressure backaction at telecom wavelengths with a hBN drum-head mechanical resonator. The thermomechanical motion of the resonator is coupled to the optical mode of a high finesse fiber-based Fabry-Pérot microcavity in a membrane-in-the-middle configuration.  We are able to resolve the optical spring effect and optomechanical damping with a single photon coupling strength of $g_0/2\pi = \SI{1200}{\hertz}$. Our results pave the way for  tailoring the mechanical properties of hBN resonators with light.  
	\end{abstract}

	\section*{Main text}
	 	
	Part of the current research efforts in the field of cavity optomechanics\cite{aspelmeyer2014} focuses on implementing nanomechanical resonators based on low dimensional materials, such as carbon nanotubes\cite{favero2008,stapfner2013,barnard2019,blien2020}, nanowires\cite{fogliano2021} or van der Waals materials\cite{song2014,singh2014,weber2016,williamson2016}. Their  low mass makes them very sensitive and responsive to external stimuli, and their large zero-point fluctuations $x_\mathrm{zpf}$ provide large optomechanical single-photon couplings $g_0$, necessary to manipulate the mechanical or optical states in the quantum regime\cite{brennecke2008,leijssen2017,peterson2019}.

	Hexagonal boron nitride (hBN) has recently caught the attention of the optomechanics community. Its  large in-plane Young's modulus of \SI{392}{\giga\pascal}\cite{zheng2017} and breaking strain of \SI{12.5}{\percent}\cite{falin2017}, together with the recent development of patterning methods\cite{elbadawi2016,kim2018} have opened the door to the engineering of mechanical resonators with high quality factors and tunable frequencies. This layered crystal is transparent in the visible and infrared part of the optical spectrum due to its wide bandgap of \SI{6}{\electronvolt}\cite{cassabois2016}, and therefore is less prone to photothermal heating than other van der Waals materials like graphene. So far, photothermal forces, rather than radiation pressure, were responsible for the optomechanical backaction observed in other two-dimensional resonators in the optical regime, limiting their performance\cite{barton2012,meyer2016, morell2019}.  
	
	Moreover, and perhaps one of the most appealing characteristics of hexagonal boron nitride, are its large variety of single-photon emitters. They span from the UV to the low infrared\cite{ kubanek2022,aharonovich2022}, are tunable via strain or electric fields\cite{grosso2017,li2020a,noh2018,nikolay2019}, and are capable of operating at room temperature and up to \SI{800}{\kelvin}\cite{kianinia2017}. Experimental works over the past three years\cite{gottscholl2020,gao2021a,stern2022} have demonstrated optically detected magnetic resonance in negatively charged boron vacancy defects ($V_B^-$), together with coherent control of the spins\cite{gottscholl2021a,gao2021}. A recent study on these vacancies has shown control at room temperature of a protected qubit basis, with a coherence time as high as \SI{0.8}{\micro\second}\cite{ramsay2023}. These spin defects are also sensitive to strain\cite{gottscholl2021,lyu2022}, making hBN a promising material to develop the field of spin-mechanics and spin-optomechanics\cite{abdi2017,wang2020}. Indeed, these hybrid systems could enable the communication between photons and qubits within hBN, a new step toward quantum networks and quantum communication, and an alternative system to the leading platform of nitrogen vacancy centers in diamond.

	Given these remarkable properties, it is natural to study how to use radiation pressure to control and modify the mechanical properties of hBN resonators. However, optomechanical dynamical backaction has not yet been observed with resonators made of this van der Waals material. Indeed, the attempts to perform cavity optomechanical experiments with hBN resonators are scarce: the first available platforms for hBN cavity optomechanics experiments are mechanically exfoliated hBN resonators coupled to the near-field of a microdisk cavity\cite{shandilya2019,liu2021}. Shandilya et al.\cite{shandilya2019} were able to measure the thermal motion of a hBN beam through its optomechanical interaction with a silicon microdisk cavity in the telecom regime for the first time, with a sensitivity of \SI{0.16}{\pm\per\sqrt{\hertz}}. Nevertheless, no  optomechanical backaction was observed.

	In this paper we report the first demonstration of radiation pressure backaction with a hexagonal boron nitride circular  drum-head resonator at room temperature. We achieve this by inserting the nanomechanical resonator in a high finesse fiber-based Fabry-Pérot microcavity\cite{rochau2021,hunger2010a,flowers-jacobs2012,pfeifer2022} in a membrane-in-the-middle configuration \cite{thompson2008,jayich2008,sankey2010,biancofiore2011, flowers-jacobs2012,aspelmeyer2014}, featuring an optical mode waist 8 times smaller than the drum diameter. The  corresponding small mode volume, intrinsic to fiber cavities, is critical to avoid clipping of the cavity mode by the edges of the drum, maintaining the high finesse of the cavity. It allows reducing the resonator's diameter down to some tens of microns, which is crucial as large area few-layer flakes are difficult to obtain through mechanical exfoliation.
	
\begin{figure}[h]
		\centering
		\includegraphics{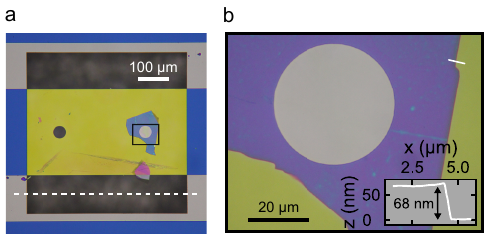}
		\caption{\textbf{hBN circular membrane.} (a) Bright field microscopy image of the suspended flake (blue) atop a \SI{40}{\micro\meter}-diameter hole patterned on a stripe processed from a commercial SiN membrane (yellow). The frame of the SiN membrane is  cleaved along the white dashed line next to the stripe to allow inserting the mechanical resonator inside the optical cavity. (b) Zoom into the black rectangle from panel (a). The inset shows	the AFM height profile along the white line in the top-right corner of the micrograph.}\label{sample}
		
	\end{figure}

	Figure~\ref{sample} shows the sample used in this work. It consists of a circular resonator made out of a  mechanically exfoliated hexagonal boron nitride flake. The flake rests on a \SI{40}{\micro\meter}-diameter circular hole patterned on a broad $500\times\SI{281}{\micro\meter\squared}$ stripe. The hole and the stripe are etched from a commercially available low stress \SI{200}{\nano\meter}-thick SiN membrane (Norcada) via standard photolithography and reactive ion etching. The crystal is transferred onto the hole through the all-dry viscoelastic method\cite{castellanos-gomez2014} and cleaned with an O\textsubscript{2}-plasma (\SI{300}{\watt}, 100\,sccm, \SI{90}{\pascal}, \SI{3}{\minute}). Its thickness is measured with an atomic force microscope in tapping mode, revealing a thickness of \SI{68}{\nano\meter} (Fig.~\ref{sample}b, inset). Finally, the \SI{200}{\micro\meter}-thick silicon frame  of the SiN membrane is cleaved  along the white dashed line in Fig.~\ref{sample}a to allow inserting the sample into the fiber-based Fabry-Pérot microcavity.
	
	\begin{figure}[h]
		\centering
		\includegraphics{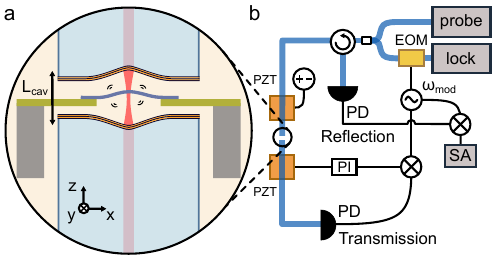}
		\caption{\textbf{Representation of the experimental setup.} (a) The Fabry-Pérot cavity is formed by two fiber mirrors (light blue rods) separated by a length of $L_\mathrm{cav}=\SI{41.8}{\micro\meter}$. The suspended hBN (dark blue) on the SiN membrane (SiN in green, Si frame in gray) is inserted between the mirrors by a positioner stack allowing tilt corrections and  position scans along the \textit{x}, \textit{y} and \textit{z} directions. (b) The cavity length is fixed by applying a DC voltage to the top piezo (PZT). The lock laser is phase-modulated at $\omega_\mathrm{mod}$ with an electro-optic modulator (EOM) and its transmission signal is demodulated at $\omega_\mathrm{mod}$ and fed to a PI controller. The latter sends the feedback signal to the bottom piezo stabilizing the cavity length. The reflection  photodetector (PD)  signal is sent to a spectrum analyzer (SA) to characterize the mechanical spectra. An additional tunable probe laser is used to measure optomechanical dynamical backaction effects.}\label{setup}
		
	\end{figure}
	 
	 	We identify the mechanical resonances by systematic characterization in a standard Michelson interferometer. Following the work from Jaeger et al.\cite{jaeger2023}, we use the experimentally determined effective mass to characterize the degree of mechanical hybridization of the hBN resonator with the mechanical modes of the SiN stripe supporting the flake (Supporting Information). We observe a series of mechanical modes between \SI{1500}{\kilo\hertz} and \SI{1700}{\kilo\hertz} that are localized in the circular hBN resonator and possess effective masses more than one order of magnitude below that of pure SiN modes, which we identify as hybridized modes with a strong hBN character. 
	 	Subsequently, the sample is placed in a high finesse Fabry-Pérot microcavity, operated at $\lambda = \SI{1550}{\nano\meter}$. The cavity, illustrated in Figure~\ref{setup}a, consists of two fiber mirrors concavely shaped by CO\textsubscript{2}-laser ablation with a radius of curvature of \SI{190.8}{\micro\meter} for  the input fiber (single mode) and \SI{135.6}{\micro\meter} for the output fiber (multi mode), respectively, separated by a length of $L_\mathrm{cav}=\SI{41.8}{\micro\meter}$. The calculated beam waist is $w_0=\SI{5.1}{\micro\meter}$. The fiber ends are ion beam sputtered with a distributed Bragg reflector (LaserOptik GmbH) to obtain a mirror transmission of 10\,ppm around $\lambda=\SI{1550}{\nano\meter}$.  Each of the fiber mirrors is glued to a shear piezo that allows tuning the cavity length  by $\pm\SI{0.7}{\micro\meter}$. By scanning the cavity length, we extract the cavity linewidth $\kappa/2\pi=\SI{18.5}{\mega\hertz}$ and the  polarization mode splitting $\Delta\nu_\mathrm{pol}=\SI{133.5}{\mega\hertz} $. The free-spectral range and empty cavity finesse are $\omega_\mathrm{FSR}/2\pi=\SI{3.584}{\tera\hertz}$ and $\mathcal{F}=194\,000$, respectively. More details on the cavity characterization methods can be found elsewhere\cite{rochau2021}. The sample is placed in a 5-axis positioning system (SmarAct) that allows aligning the sample to the cavity mode axis (z-axis in Figure~\ref{setup}a) and scanning it along the \textit{x}, \textit{y} and \textit{z} directions. The system is operated at room temperature, in vacuum at a pressure below 10\textsuperscript{-6}\,mbar and inside an acoustic isolation box to damp spurious mechanical vibrations. Figure~\ref{setup}b depicts a simplified version of the setup. To perform dynamical backaction experiments, we lock the cavity length to a reference laser (NKT Koheras Basik E15, $\lambda = \SI{1550}{\nano\meter}$) referred in the following as lock laser. The lock laser is  phase-modulated at $\omega_\mathrm{mod}/2\pi=\SI{30}{\mega\hertz}$  with an electro-optic modulator (EOM). The light transmitted through the cavity, recorded by a fast photodetector (PD), is demodulated at the same frequency and  sent to a PI controller that sends the feedback signal to  a piezo (PZT) to perform the lock. We use the Y-quadrature of the demodulated signal as error for the lock. The  reflected light from the lock laser is equally demodulated and fed to a spectrum analyzer that records the mechanical spectra. We use an additional probe laser (New Focus TLB-6700) to sweep the wavelength across the cavity resonance and exert an optomechanical force. Both lasers are operated at orthogonal cavity polarizations to avoid interference effects.

	\begin{figure*}[h]
		\centering
		\includegraphics{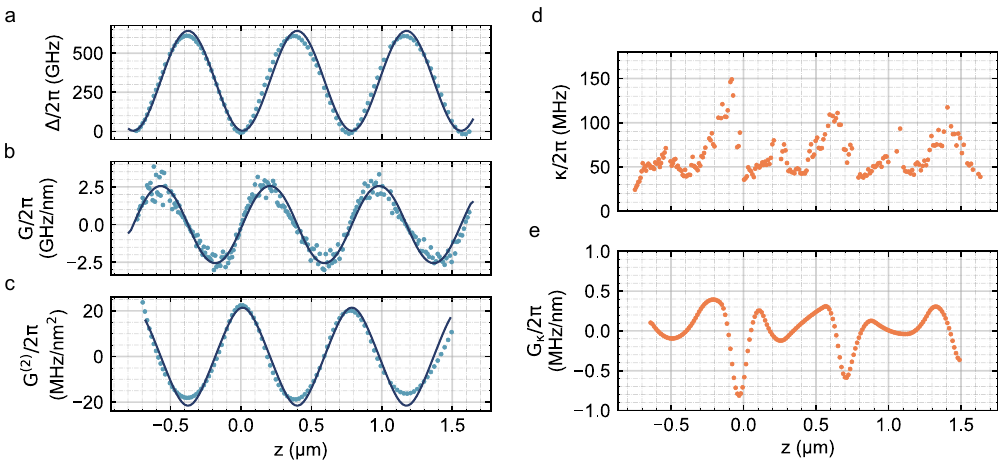}
		\caption{\textbf{Static optomechanical couplings.} (a) Cavity detuning (dots) versus sample position \textit{z} along the cavity mode axis. The coordinate $z = 0$ corresponds to the sample sitting in an optical field node. (b) Dispersive and (c)  quadratic dispersive coupling . The solid lines in  (a), (b) and (c) are the result of transfer matrix calculations. (d) Cavity linewidth modulation and (e) corresponding dissipative coupling.}\label{staticspec}
		
	\end{figure*}
	The dispersive optomechanical coupling $G=\partial\omega_\mathrm{cav}/\partial z$\cite{biancofiore2011} is obtained by measuring the dependence of the detuning of the cavity resonance frequency $\Delta(z)$ as a function of the sample position \textit{z} along the cavity mode axis. To that end, the system is operated in an open-loop configuration and both piezos are scanned symmetrically. We use only the lock laser and record the cavity transmission and reflection. We observe a sinusoidal pattern of the detuning (blue dots, Fig.~\ref{staticspec}a)  that reaches a maximum of $\Delta/2\pi = \SI{600}{\giga\hertz}$ when the sample is placed at the cavity antinodes. We verify that the cavity detuning is originated solely from the hBN drum and has no contribution from the surrounding SiN by performing the same measurement on the second, empty \SI{40}{\micro\meter}-diameter hole  (Supporting Information) shown in Figure~\ref{sample}a. $G$  is directly extracted from the measurement by performing a numerical derivative (blue dots, Fig.~\ref{staticspec}b). We observe a maximum of $\rvert G/2\pi\lvert=\SI{2.5}{\giga\hertz\per\nano\meter}$. As expected for a membrane-in-the-middle system, the quadratic dispersive coupling $G^{(2)} = \partial^2\omega_\mathrm{cav}/\partial z^2$ reaches a maximum at the cavity nodes and antinodes\cite{sankey2010,karuza2012}, of  value $\rvert G^{(2)}/2\pi\lvert=\SI{20}{\mega\hertz\per\nano\meter\squared}$ for our system (blue dots, Fig.~\ref{staticspec}c). The quadratic coupling is extracted by smoothing the data from Fig.~\ref{staticspec}b with a spline and computing its numerical derivative. The measurements are well reproduced by transfer matrix calculations \cite{katsidis2002}  (solid lines in Figure~\ref{staticspec}), with input parameters a flake thickness  of \SI{58}{\nano\meter} and flake refractive index of $n_\mathrm{hBN} = 1.85$. The reduced thickness originates from an additional O\textsubscript{2} cleaning step that etched the flake, which was observed as a change of color under the optical microscope.  
	
	Any absorption or scattering from the flake will manifest as a modulation of the cavity linewidth $\kappa$, displayed in Figure~\ref{staticspec}d. The linewidth is calculated from the cavity transmission on resonance (see Ref.~\cite{rochau2021} for more details), and reaches a minimum value of $\kappa/2\pi \simeq \qty{40}{\mega\hertz}$ when the sample is placed at the cavity nodes.  hBN has a negligible absorption coefficient at telecom wavelengths\cite{rah2019,lee2019}, and therefore the large cavity linewidths we observe cannot be attributed to optical absorption. Performing the same measurements shown in Fig.~\ref{staticspec}d on the SiN membrane results in modulations of the same magnitude (Supporting Information) that cannot be explained considering the absorption coefficient of SiN. This suggests that the losses in our system are dominated by scattering to higher order optical modes of the cavity originated by a sample misalignment with respect to the cavity mode axis\cite{karuza2012,sankey2010}, and not by remaining impurities from the transfer process. The hypothesis is also consistent with the dips observed at the antinodes and the small scale variations ($z<\lambda$) of the cavity linewidth\cite{sankey2010}. The linewidth modulation translates into the dissipative coupling $G_\kappa=\dfrac{\partial\kappa}{\partial z}$ (Figure~\ref{staticspec}e) that we extract by smoothing the data in Fig.~\ref{staticspec}d and performing a numerical derivative. We highlight that dissipative coupling is more than three orders of magnitude smaller than the dispersive coupling and is therefore negligible in our system.

\begin{figure}[h]
		\centering
		\includegraphics{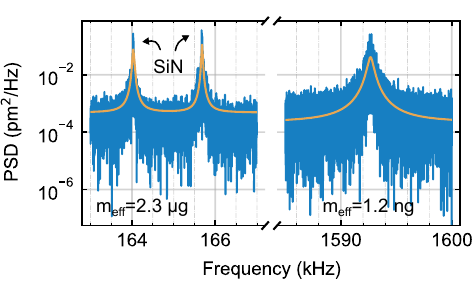}
		\caption{\textbf{Displacement power spectral density (PSD) measured with the locked cavity.} Two SiN modes are shown in the lower frequency range. The left one at \SI{164.0}{\kilo\hertz}, of known effective mass $m_\mathrm{eff} = \SI{2.3}{\micro\gram}$, is used to calibrate the spectra. The higher frequency range shows the hBN mode  used for dynamical backaction experiments. Solid lines are theoretical fits using Eq.~\ref{sx} with an additional constant noise floor. }\label{calib}
		
	\end{figure}

	To perform dynamical backaction experiments, we place the hBN drum  $\Delta z = \SI{13}{\nano\meter}$ away from the cavity node. From the cavity transmission, we obtain a loaded cavity linewidth of $\kappa/2\pi=\qty{39.8(2)}{\mega\hertz}$, yielding a loaded cavity finesse of $\mathcal{F} = 90\,000$. This sample position corresponds to a value of $G/2\pi = \SI{330}{\mega\hertz\per\nano\meter}$ and  $G_{\kappa}/2\pi = \SI{0.19}{\mega\hertz\per\nano\meter}$, extracted from the measurements in Fig.~\ref{staticspec}.  We lock the cavity length with a lock power of $P_\mathrm{l} = \SI{34.5}{\micro\watt}$ and at a frequency $\omega_\mathrm{cav}$ detuned from the lock laser frequency $\omega_\mathrm{l}$ by $\Delta_\mathrm{l}/2\pi= (\omega_\mathrm{l}-\omega_\mathrm{cav})/2\pi = -\SI{50}{\mega\hertz}$. We choose this value as a compromise between having a good signal in reflection and minimizing the backaction from the lock. To transform the measured power spectral densities $S_\mathrm{V}$ in volts into displacement power spectra densities $S_\mathrm{x}$, we use as reference the thermomechanical spectra of a well characterized SiN mode at \SI{164.0}{\kilo\hertz} of known effective mass $m_\mathrm{eff} = \SI{2.3}{\micro\gram}$ when probed at the center of the drum-head, following the calibration procedure described in Ref.~\citenum{hauer2013}. This mode is suitable for the calibration because it does not show optomechanical coupling. The conversion factor between displacement and volts is \qty{3}{\pico\meter\per\milli\volt}.  Figure~\ref{calib} shows the spectrum of the SiN mode used for calibration, together with the spectrum of the hBN mode used to perform backaction experiments. The solid lines are fits using the displacement power spectral density formula
	\begin{equation}\label{sx}
		S_\mathrm{x}(\omega) = \dfrac{4\Gamma_0 k_B T}{m_\mathrm{eff}\bigg(\big(\Omega_0^2-\omega^2\big)^2+\Gamma_0^2\omega^2\bigg)},
	\end{equation}
	where $T$ is the bath temperature, $k_B$ the Boltzmann constant,  and where the only fit parameters are the mechanical resonance frequency $\Omega_0$, the mechanical linewidth $\Gamma_0$, the effective mass $m_\mathrm{eff}$ and an additional constant noise floor. The fit yields, assuming the sample is thermalized with the environment,  $\mathrm{\Omega_0/2\pi=\SI{1592.671\,(4)}{\kilo\hertz}}$,  $\mathrm{\Gamma_0/2\pi=\SI{525\,(10)}{\hertz}}$,  quality factor $\mathrm{Q=3000\,(50)}$ and $m_\mathrm{eff}=\SI{1.2\,(1)}{\nano\gram}$ for the hBN mode. The inferred zero-point fluctuations are  $x_\mathrm{zpf}=\sqrt{\hbar/(2m_\mathrm{eff}\Omega_0)}=\SI{2.1\,(1)}{\femto\meter}$ which lead to a single-photon coupling strength of $g_0/2\pi = G x_\mathrm{zpf}/2\pi = \SI{330}{\mega\hertz\per\nano\meter}\cdot\SI{2.1}{\femto\meter}=\SI{693}{\hertz}$.  Since the effective mass was measured with the lock red detuned, the calculated value of $g_0$ is a lower limit to the real value. Finally, from the noise floor level we obtain a sensitivity of \SI{15}{\femto\meter\per\sqrt{\hertz}}.

\begin{figure}[h]
		\centering
		\includegraphics{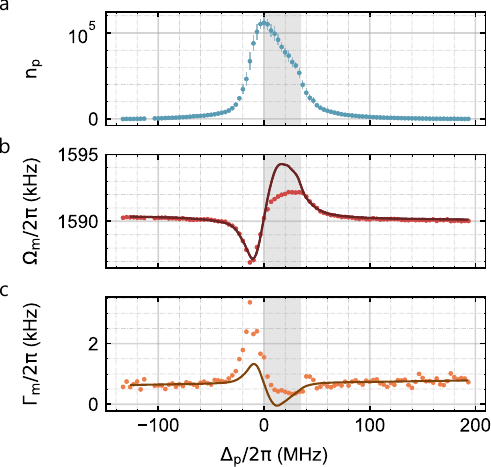}
		\caption{\textbf{Radiation pressure backaction.} (a) Experimental probe photon number $n_\mathrm{p}$ versus probe detuning $\Delta_\mathrm{p}$. (b) Optical spring effect (red dots) and  theoretical regression (solid line). The gray region corresponds to the regime where the photon number shows a nonlinear response and its data are excluded in the fit. (c) Mechanical damping (orange dots) and theory calculation  (solid line) with the parameters from the fit in (b). The errorbars  for $\Omega_m$ and $\Gamma_m$, generated from the fits to Eq.~\ref{sx}, are omitted because they are not appreciable in the figure. }\label{dynback}
		
	\end{figure}

	  The mechanical resonator will experience an optomechanical force when the frequency of the probe laser $\mathrm{\omega_\mathrm{p}}$  is swept  across the cavity resonance, with $\mathrm{\Delta_\mathrm{p} = \omega_\mathrm{p} - \omega_\mathrm{cav}}$ the probe detuning. This leads to the optical spring effect, a shift of the mechanical frequency $\Omega_\mathrm{m}^2 = \Omega_0^2 + \delta(\Omega^2)$ that, when neglecting the dissipative coupling, is given by\cite{biancofiore2011}
	 
	 \begin{equation}\label{omega}
	 	\delta(\Omega^2) = 2\Omega_0 g^2\bigg(\dfrac{\Omega_0 + \Delta_\mathrm{p}}{(\Omega_0 +\Delta_\mathrm{p})^2+(\kappa/2)^2}-\dfrac{\Omega_0-\Delta_\mathrm{p}}{(\Omega_0-\Delta_\mathrm{p})^2+(\kappa/2)^2}\bigg),
	 \end{equation}
 	 where $g = \sqrt{n_\mathrm{p}}g_0$ is the dispersive optomechanical coupling strength, $n_\mathrm{p}$ the circulating probe photon number inside the cavity and $g_0 = G x_\mathrm{zpf}$ the single-photon coupling strength. The mechanical linewidth experiences a shift as well $\Gamma_\mathrm{m} = \Gamma_0 + \Gamma_\mathrm{opt}$, with $\Gamma_\mathrm{opt}$ the optomechanical damping. Considering negligible dissipative coupling, the latter reads\cite{biancofiore2011}
 	 
 	 \begin{equation}\label{gamma}
 	 	\Gamma_\mathrm{opt} = 2g^2\kappa\bigg(\dfrac{1}{(\Omega_0+\Delta_\mathrm{p})^2+(\kappa/2)^2}-\dfrac{1}{(\Omega_0-\Delta_\mathrm{p})^2+(\kappa/2)^2}\bigg).
 	 \end{equation}
	 In Figure~\ref{dynback} we demonstrate dynamical backaction by measuring the optical spring effect (Figure~\ref{dynback}b) and optomechanical damping (Figure~\ref{dynback}c) of the hBN mechanical mode in Fig.~\ref{calib}. We scan the probe frequency across the cavity resonance  using a constant probe power of $P_\mathrm{p} = \SI{11}{\micro\watt}$. For each probe detuning, we extract  the circulating probe photon number  $n_\mathrm{p}$ inside the cavity from its transmission and measure the thermomechanical spectrum. We obtain  the effective mechanical resonance frequency $\Omega_\mathrm{m}$ and effective linewidth $\Gamma_\mathrm{m}$ by fitting Eq.~\ref{sx} to the spectra. The circulating probe photon number (Fig.~\ref{dynback}a) appears to be nonlinear and is asymmetric with respect to the probe detuning. Starting from positive detunings, the photon number increases in a Lorentzian-fashion until $\Delta_\mathrm{p}/2\pi =\qty{35}{\mega\hertz}$. Then, it increases almost linearly until reaching its maximum at $\Delta_\mathrm{p}/2\pi =0$. After that, it drops with decreasing detuning according to a Lorentzian trend.   This behavior indicates the presence of a nonlinearity in the system that we will discuss in the next paragraphs.
		
	The backaction of the probe leads to a softening  (stiffening) of the mechanical frequency for negative (positive) detunings (Fig.~\ref{dynback}b, red dots), with a maximum frequency shift of $\rvert\delta\Omega/2\pi\lvert = \SI{3}{\kilo\hertz}$ for negative detunings. In addition, the natural frequency has a slight negative slope stemming from drifts in the lock that originate from changes of the sample position due to slow temperature drifts in the laboratory.  The solid line is a  fit to the optical spring using Eq.~\ref{omega} together with a linear background. For the fit, we exclude the data points where the photon number shows a nonlinear response ($0<\Delta_\mathrm{p}/2\pi<\qty{35}{\mega\hertz}$, gray shaded area in Fig.~\ref{dynback}b and Fig.~\ref{dynback}c). We also assume negligible dissipative coupling, and feed the equation with the experimentally measured probe photon number $n_\mathrm{p}$.  The only free parameters for the fit are $g_0$ and the cavity linewidth $\kappa$. The best fit yields  $g_0/2\pi=\SI{1200\,(20)}{\hertz}$ and $\kappa/2\pi = \qty{54\,(1)}{\mega\hertz}$.  The data shows an asymmetry around the probe detuning that is not predicted by the linear cavity optomechanics theory (Eqs.~\ref{omega} and \ref{gamma}).   We discard the possibility of an asymmetry induced by the dissipative coupling because of the small value of the single-photon dissipative coupling strength, $g_{0,\kappa}/2\pi = G_{\kappa}x_\mathrm{zpf}/2\pi= \SI{0.4}{\hertz}$. The asymmetry in the optical spring could be attributed to nonlinearities in the system given the large photon number  reached in the experiments ($n_\mathrm{p}>10^5$). Indeed, a cavity optomechanical system incorporating a negative Kerr nonlinearity can reproduce the asymmetries we observe\cite{zoepfl2023}. Additional experiments should be conducted in the future to clarify the origin of the nonlinearity in the system. This could be due to, for example, the radiation pressure force itself \cite{dorsel1983} or to thermal effects given the absorption of the mirrors' coating \cite{hunger2010a}. 
	
	Figure~\ref{dynback}c shows the effective linewidth $\Gamma_\mathrm{m}$ (orange dots) as a function of the probe detuning. The solid line is the result of Eq.~\ref{gamma} using  the parameters obtained from the fit of the optical spring. We observe a larger broadening (optomechanical cooling) than what is predicted by the theory for negative detunings, whereas the narrowing (optomechanical heating) found for positive detunings is less pronounced compared to the theoretical model. This behavior also matches the one of a cavity optomechanical system with a negative Kerr nonlinearity \cite{zoepfl2023}, which supports the hypothesis of a nonlinearity present in the system. In addition, the measured linewidths are broadened due to unavoidable mechanical  fluctuations of the positioning system and the fiber mirrors. The fluctuations translate into cavity length noise, which directly affects the detuning and therefore turn into mechanical frequency noise \cite{flowers-jacobs2012,rochau2021}. Because each experimental mechanical spectrum is the average of the mechanical response during the acquisition time of the spectrum analyzer, the experimental resonance frequency matches on average the theoretical value, besides the asymmetry already discussed. The mechanical linewidth (Fig.~\ref{dynback}c), however, is  broadened especially for detunings at which $\Omega_\mathrm{m}$ depends strongly on the detuning. The positioning system's mechanical fluctuations are also the limiting factor of this experiment and the reason why we are unable to successfully lock the cavity at sample positions with larger coupling $G$.

	
	To conclude, we have demonstrated radiation pressure backaction on a hBN resonator at telecom wavelengths. The mechanical performance of our sample was limited by the hybridization to the modes of the low stress SiN membrane resonator and the mechanical imperfections originating from the dry transfer process. The former can be improved by using high stress Si\textsubscript{3}N\textsubscript{4} as a support for the hBN drums, pushing the Si\textsubscript{3}N\textsubscript{4} resonances to higher frequencies and allowing resolution of the distinct mode shapes; the latter by using more gentle transfer mechanisms like a wet transfer\cite{jaeger2023}. The setup presented in this work is very versatile: a careful design of the mirror coatings would enable the incorporation of several wavelengths,  enabling photoluminiscence experiments on-site. This will open the door to new ways of studying the strain dependence of hBN defects with an additional control of the mechanical properties through light, and a new step forward toward the realization of spin-optomechanics with this van der Waals material.
	\vspace{1.5cm}

	\section*{Acknowledgments}
	We are grateful to Avishek Chowdhury, Francesco Fogliano and Felix Rochau for useful discussions. We also thank David Hunger for his assistance in the fabrication of the fiber mirrors. I.S.A. acknowledges support from the European Union’s Horizon 2020 research and innovation program under the Marie Sklodowska-Curie Grant Agreement No. 722923 (OMT). I.S.A. and E.M.W. acknowledge the Bavarian State Ministry of Science and Arts via the project EQAP. K.W. and T.T. acknowledge support from JSPS KAKENHI (Grant Numbers 19H05790, 20H00354 and 21H05233).
	
	\section*{Author contributions}
	I.S.A. fabricated the samples, constructed the setup and carried out all the experiments. I.S.A. and E.M.W. analyzed and interpreted the data. T.T. and K.W. grew the hBN crystals. I.S.A. and E.M.W. contributed to writing the manuscript.

	\newpage
	\providecommand{\latin}[1]{#1}
\makeatletter
\providecommand{\doi}
  {\begingroup\let\do\@makeother\dospecials
  \catcode`\{=1 \catcode`\}=2 \doi@aux}
\providecommand{\doi@aux}[1]{\endgroup\texttt{#1}}
\makeatother
\providecommand*\mcitethebibliography{\thebibliography}
\csname @ifundefined\endcsname{endmcitethebibliography}
  {\let\endmcitethebibliography\endthebibliography}{}

	

\section*{Supplementary material}

# Supporting Information Available

## S1  Sample bulging

Figure S1a shows a reflection micrograph of the sample after cleaving the bottom part of the Si frame of the membrane. During the cleaving process, the SiN membrane (yellow) got damaged and a small part broke off, but the flake remains intact. Figure S1b is the same as Figure S1a but with enhanced phase contrast, revealing a wrinkle more clearly than what is appreciable in the main text. Additionally, the flake surface has a large phase contrast probably due to bulging or uneven tensioning. This effect, that may originate from the exfoliation and/or transfer process, has already been observed in other mechanically exfoliated resonators.[1,2]

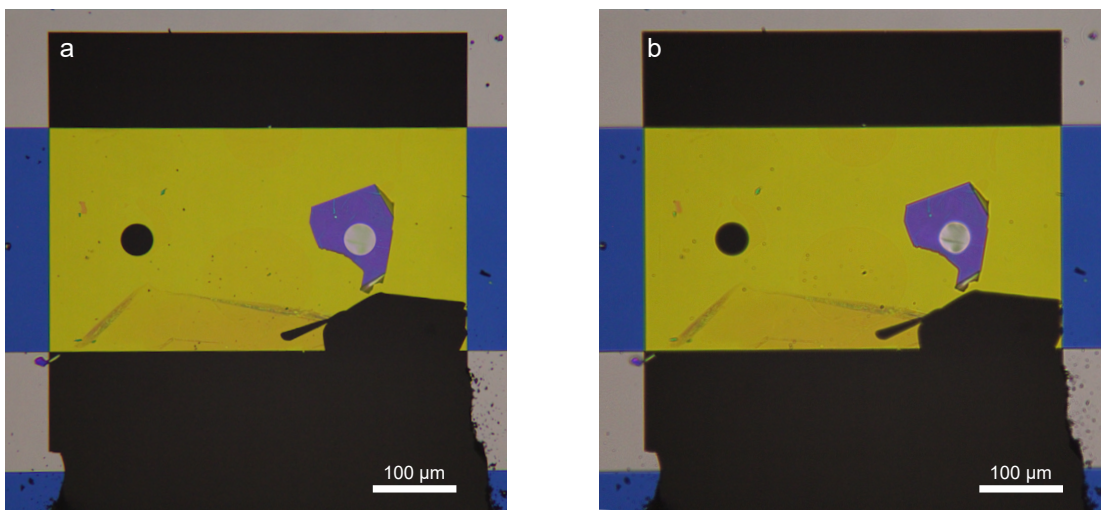

Figure S1: **Cleaved sample.** (a) Bright field microscopy image. (b) Same settings as in (a) but with increased phase contrast obtained by closing enough the reflection microscope's condenser aperture.

## S2  Losses from the circular frame

Figure S2a displays a micrograph of a sample used for investigating the clipping and scattering losses introduced by the circular frame of the drums. It consists of a low-stress 200-nm

S-2

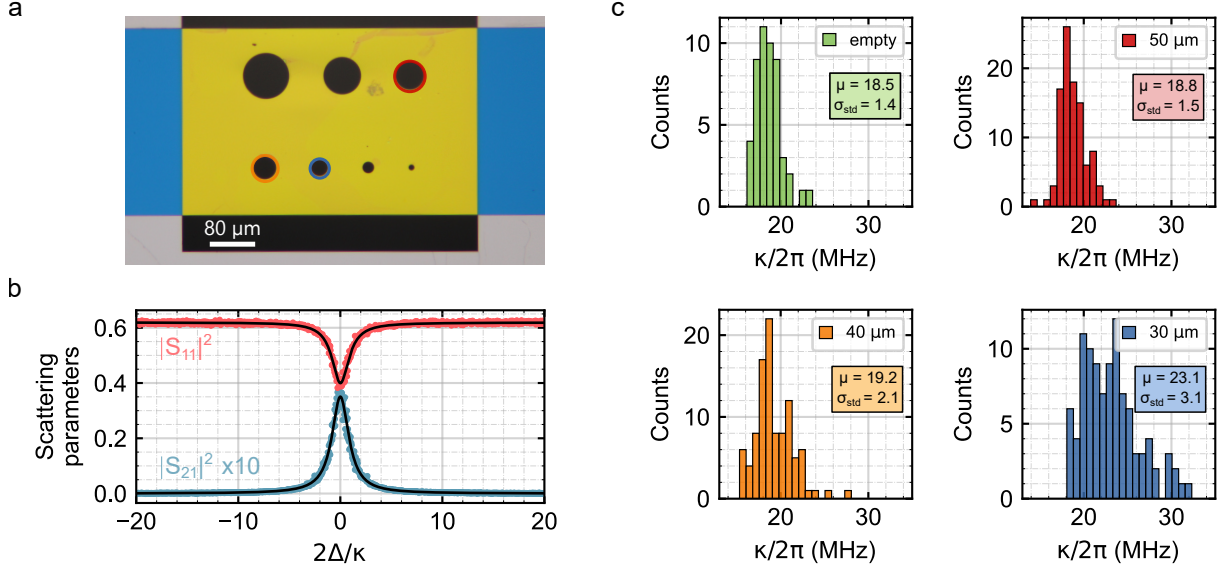

Figure S2: **Losses from the circular frame.** (a) Microscope image of the sample used for the test. The holes used for the measurement have a false-colored rim according to the legend in (c). (b) Empty cavity scattering parameters as a function of normalized detuning. Reflection $|S_{11}|^2$ (red dots) and transmission $|S_{21}|^2$ (blue dots, amplified by a factor of 10), together with the fits (black solid lines) used to extract the cavity linewidth. (c) Histogram of cavity linewidth for an empty cavity (green) and for different hole diameters (red, orange, blue) specified in the legend. The graphs include the mean value $\mu$ and standard deviation $\sigma_{\text{std}}$ of the histogram in units of MHz.

thick SiN stripe with seven etched holes of different diameters. The latter are, from largest to smallest, 80 µm, 65 µm, 50 µm, 40 µm, 30 µm, 20 µm and 10 µm. In each measurement, the hole is centered with respect to the cavity mode axis. Then, the sample position along the cavity mode axis is changed 100 times in a $\Delta z = 1$ µm range. At each position we record the cavity transmission and reflection.

For reference, Figure S2b shows an example of the cavity transmission and reflection when it is empty (no sample). The fits used to extract the cavity linewidth are based on the model from Gallego et al.,[3] which accounts for the mode-filtering effect intrinsic to fiber cavities. Figure S2c displays histograms of the cavity linewidth for the empty case (green) and for hole diameters of 50 µm (red), 40 µm (orange) and 30 µm (blue). For diameters $\leq 20$ µm the losses are too large and the cavity mode is lost, so consequently there is no data to show. The mean $\mu$ and standard deviation $\sigma_{\text{std}}$ of the linewidth are displayed in the

graphs. The standard deviation from the hole with a diameter of 50 µm corresponds to that of the empty cavity. It increases by a factor of 1.3 for the 40 µm-hole, but the mean value remains within the error of the measurement. However, $\sigma_{\text{std}}$ doubles the empty cavity value for the 30 µm-hole, and the mean is displaced towards larger linewidths. In particular, the linewidth is increased up to 32 MHz. This corresponds to an additional round trip loss of $\mathcal{L}_{\text{rt}} = \mathcal{L}_{\text{clip}} + \mathcal{L}_{\text{scat}} = 20$ ppm. We estimate an upper bound to the beam waist of the cavity using $\mathcal{L}_{\text{rt}}/2 = \exp(-D^2/(2w_0^2))$,[4] where $D = 30$ µm is the hole diameter and where we have assumed $\mathcal{L}_{\text{scat}} = 0$. This yields $w_0 = 6.3$ µm.

## S3 Mechanical mode identification

### S3.1 Michelson interferometer

The mechanical properties of the sample are characterized in a standard Michelson interferometer, operated at room temperature and equipped with a He-Ne laser ($\lambda = 632$ nm, 180 µW). The sample is mounted on a 3-axis positioning system (Attocube ECSx3030) and placed inside of a vacuum chamber to reduce gas damping ($< 10^{-4}$ mbar). The interference from the light reflected from the sample and the reference arm is detected via a fast photodetector and sent to a spectrum analyzer. We actively stabilize the signal to the middle of the Michelson interferometer fringe by mounting the reference arm mirror on a piezo element controlled with a proportional-integral feedback loop. We send to the piezo element a ramp signal of peak-to-peak voltage $V_{\text{pp}} = 10$ V that allows a 3-bright-fringes scan of fringe amplitude $V_{\text{f}} = 180$ mV. After stabilization, small voltages $\delta V(t)$ ($\delta V(t) \ll V_{\text{f}}$) are converted into displacement via $\delta x(t) = \delta V(t)\lambda/(4\pi V_{\text{f}})$, which allows a direct conversion of the measured power spectral density (PSD) in volts to displacement. We fit Eq. 1 from the main text to the displacement power spectral density and extract the effective mass $m_{\text{eff}}$, resonance frequency $\Omega_0$, linewidth $\Gamma_0$ and quality factor $Q = \Omega_0/\Gamma_0$ of detected mechanical modes.

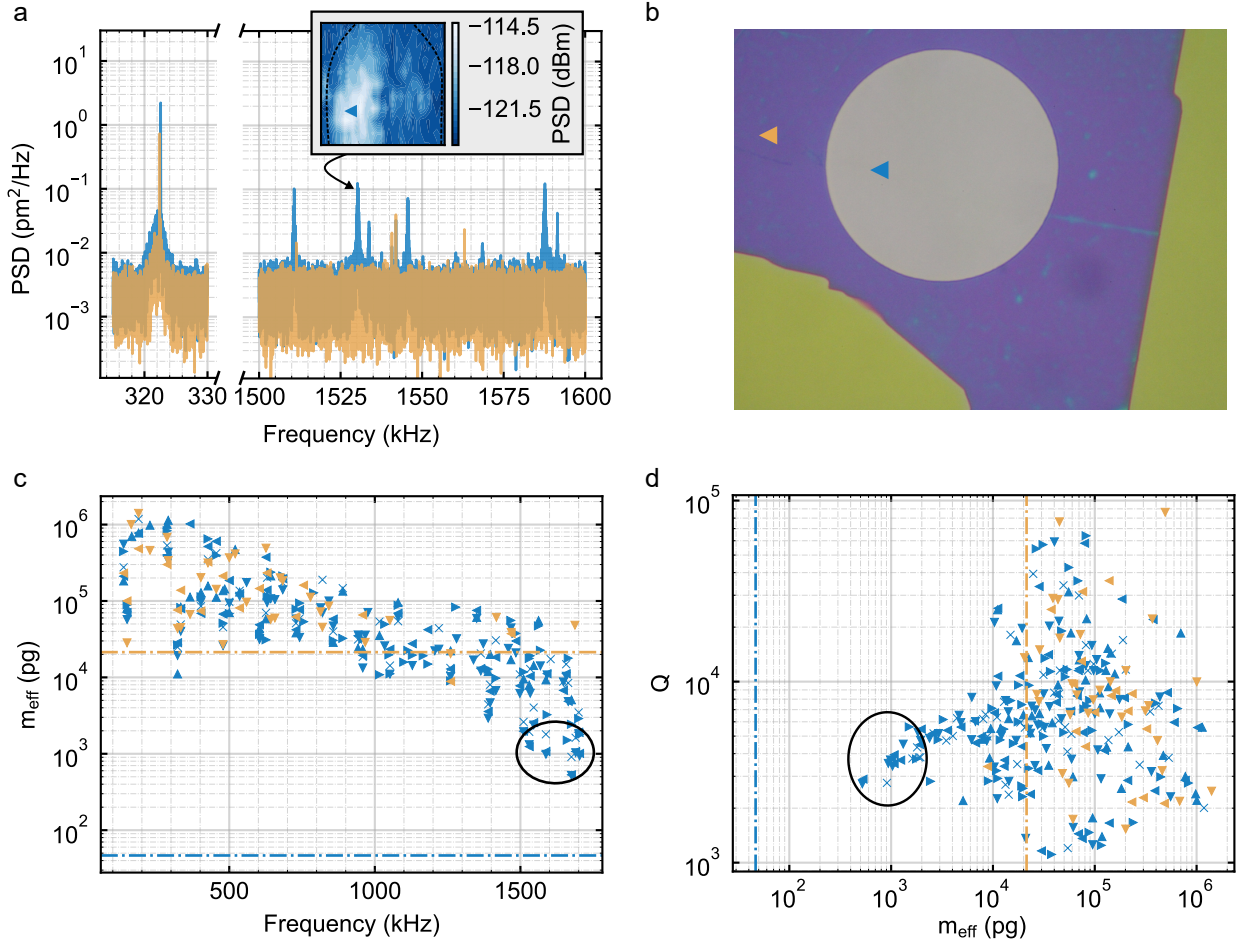

Figure S3: **Mechanical characterization**. (a) Power spectral density (PSD) on the suspended flake (blue line, measured at the blue triangle in the inset and in (b)) and on the flake resting on the SiN membrane (orange line, orange triangle in (b)). The inset shows the maximum of the PSD in dBm for the 1530.3 kHz mode (black arrow) as a function of spatial position measured with a bandwidth of 1 Hz. The black dashed lines represent the measured edges of the circular drum. Due to hysteresis in the positioner's scan directions the plot is distorted and not to scale. (b) Microscopy image of the 40 µm-diameter drum. The triangles depict the locations at which the measurements in (a) were taken. (c) Experimentally determined effective mass versus resonance frequency. (d) Quality factor versus effective mass. Blue markers are resonances measured on the hBN drum whereas orange markers depict resonances measured on the hBN on top of SiN. Different marker shapes indicate measurements taken at different positions. Up, down, left, and right triangles indicate measurements done in the upper, bottom, left and right side of the hBN drum. The cross depicts the center. Dashed lines represent the theoretical effective masses for ideal hBN (blue) and SiN (orange) resonators. The black ovals highlight the interval in which the dynamical backaction measurements from the main text take place.

Figure S3a displays the thermomechanical motion's power spectral density measured at

different positions: The orange and blue lines are measured on the hBN flake, outside and on the circular resonator, respectively (orange and blue left triangles in Figure S3b). Both measurements display a pronounced resonance at 322.6 kHz attributed to the SiN membrane; the wider feature next to it, at 322.0 kHz, is noise from the He-Ne laser. For frequencies above 1 MHz a plethora of peaks start arising on the hBN resonator that are not present in the spectra measured on the SiN membrane. The right part of Figure S3a shows the frequency range between 1.5 MHz and 1.6 MHz as an example. We map the mode shapes of some of these modes (inset in Fig. S3a), revealing a strong localization (see section below for more information on the mode mapping measurements). All this suggests that these modes, undetectable on the SiN membrane, originate purely from the hBN circular resonator or result from hybridization between the modes intrinsic to the hBN drum-head and the SiN membrane ones.[5,6]

To further investigate the degree of hybridization, we measure the effective mass of all modes starting from the fundamental mode of the SiN membrane at 137.2 kHz until 1.7 MHz. Any hybridization between the heavy SiN membrane and the hBN drum will result in an increase of the effective mass of the hBN resonator.[6] Fig. S3c displays the effective mass as a function of resonance frequency. The data points are obtained from fitting the response of the resonator driven by thermal noise measured at different positions. Blue markers depict measurements on the free-standing hBN resonator whereas orange ones on the flake resting on the SiN membrane. Different marker shapes indicate measurements taken at different locations. The orange line represents the theoretical effective mass of an ideal rectangular SiN membrane of dimensions $500 \times 281 \times 0.2\,\mu m^3$, $m_{\text{eff,SiN}}^{\text{th}} = 0.25 m_0 = 21.5\,\text{ng}$, with $m_0$ the physical mass, where we have adopted the effective mass definition from Hauer et al.[7] This value sets a lower limit for the effective mass of pure SiN modes in case of an ideal rectangular geometry. Similarly, using a mass density of $2100\,\text{kg}\,\text{m}^{-3}$,[8] we predict for the fundamental mode of the hBN drum $m_{\text{eff,hBN}}^{\text{th}} = 0.26 m_0 = 46.7\,\text{pg}$, shown as blue dashed line in Fig. S3c. This is the upper limit of the effective mass of pure ideal hBN drum modes. From our data

we observe that the effective masses measured outside the circular drum (orange markers) stay close to the theoretical limit for SiN, which is what we expect as no hybridization should occur. However, the modes measured on the free standing hBN resonator (blue data set) start to deviate from the SiN limit toward lower masses for frequencies above 1 MHz: this is a sign of mode hybridization. The effective mass in this case reaches a minimum value of 520 pg at 1675.2 kHz. The resonance frequency of the fundamental mode of an ideal circular drum with our sample's geometrical parameters is $\Omega_0/2\pi = 960$ kHz. Since the spectrum of modes of the SiN is very dense above 400 kHz, we expect some degree of hybridization for all hBN modes. We were not able to identify the textbook modes of a circular membrane in our sample and we attribute this to the large bulging originating from the dry transfer (Figure S1).[1,2] Nevertheless, we consider that the modes with effective masses more than an order of magnitude smaller than the limit for the SiN resonator have a very pronounced hBN drum-head character. Those are the modes we use to demonstrate dynamical backaction on the hBN resonator in the main text.

Fig. S3d shows the dependence of the quality factor with the effective mass. The modes with an effective mass below 2 ng, with a strong hBN drum-head character, have quality factors in the range $Q \in [2000, 6000]$. This is in agreement with other measured Q factors of large diameter 2D material drums,[9,10] however these Q factors may be influenced by the hybridization.

## Mode mapping

The mode shapes of some mechanical resonances are obtained by recording the spectra while scanning the sample position along the drum's surface. At each position of the scan measurements we record the PSD with the spectrum analyzer and the reflection signal with an oscilloscope. The travel range of the reference arm mirror is not long enough to cover for changes in reflection when the sample is scanned over regions of different material. For this reason, every time the scans go over a different material the lock is lost. Figure S4a and Fig-

ure S4b show the maximum of the PSD in dBm of the mechanical spectra of the 1530.3 kHz mode measured with a 1 Hz bandwidth and the standard deviation of the reflection signal, respectively, for a 32 × 11 scan. The standard deviation of the reflection signal is used to draw the edges of the resonator depicted in Fig. S3a. The travel range of the scanning mode of our positioning system is too small to cover the whole drum-head and we have to use the stepper function. The latter suffers from a strong hysteresis, and consequently the colormap's $x$ and $y$ axis are not to scale and distorted; for spatial reference, the measured edges of the drum, drawn as guidelines to the eye from the maps of the standard deviation, are depicted as black dashed lines superimposed to the colormaps. The mode shape reveals that the mechanical mode is confined within the circular resonator, with a prominent lobe on the left side. Similar mode shapes have been observed for exfoliated hBN resonators.[2] The deviation from the expectation for an ideal circular resonator[7] is attributed to the uneven tensioning and/or bulging of the resonator resulting from the transfer process.[1,2] A closer look into the sample shows indeed hints of bulging (Fig. S1).

Figure S4c and Figure S4d display the same measurements but for the SiN mode at 322.6 kHz and a 15 × 15 scan. The peak amplitude of this mode remains constant over the scanned area at $100 \pm 1$ dBm .

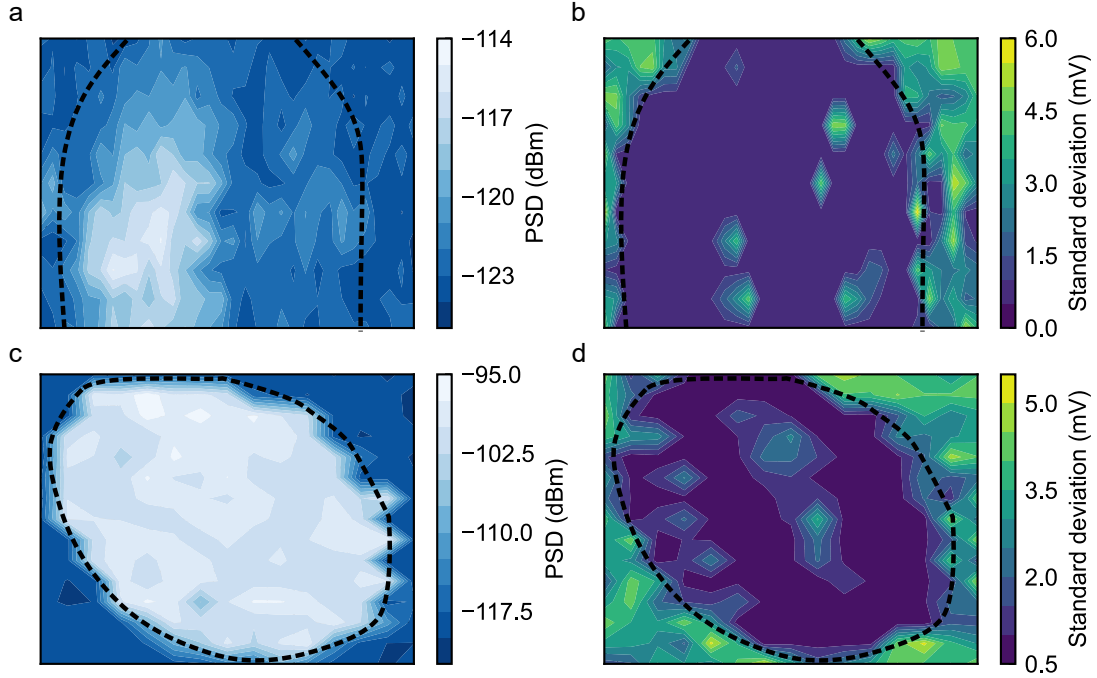

Figure S4: **Mode mapping**. (a) Amplitude of the 1530.3 kHz mode as a function of spatial position. (b) Standard deviation of the reflection signal. (c) Amplitude of the 322.6 kHz mode as a function of spatial position. (d) Standard deviation of the reflection signal for the same mode as in (c).

## S3.2 Fabry-Pérot cavity

The cavity voltage PSD is transformed into displacement PSD following the procedure described in the main text. We choose the SiN mode at 164.0 kHz to calibrate the spectra because it is not affected by dynamical backaction. Figure S5a displays a typical mechanical spectrum taken at the center of the hBN resonator after calibration. We can only record the spectra at that position (±5 µm); getting closer to the edges generates too many scattering losses and the cavity mode is lost. This, together with the fact that the cavity lock is lost when we perform a step movement with the positioning system, makes us unable to map the mode shapes inside of the cavity with the current setup capabilities. Adding a second laser operating at shorter wavelengths, outside the coating band of the mirrors, would enable standard interferometry measurements like in the Michelson setup. Nevertheless, we can extract the effective masses from the measurement for the available position (Fig. S5b).

This allows us to distinguish the modes with a strong hBN character by selecting those with effective masses below 2 ng. The mode used for the radiation pressure backaction measurements is highlighted with a dark arrow in both figures. The difference between the effective masses shown here with the ones from Fig. S3 arise from the optomechanical interaction: because the cavity is locked, the effective spring constant (or effective mass) is modified by the presence of the lock laser. This can affect different modes at different magnitudes, given by the overlap of the mechanical modes with the optical one.

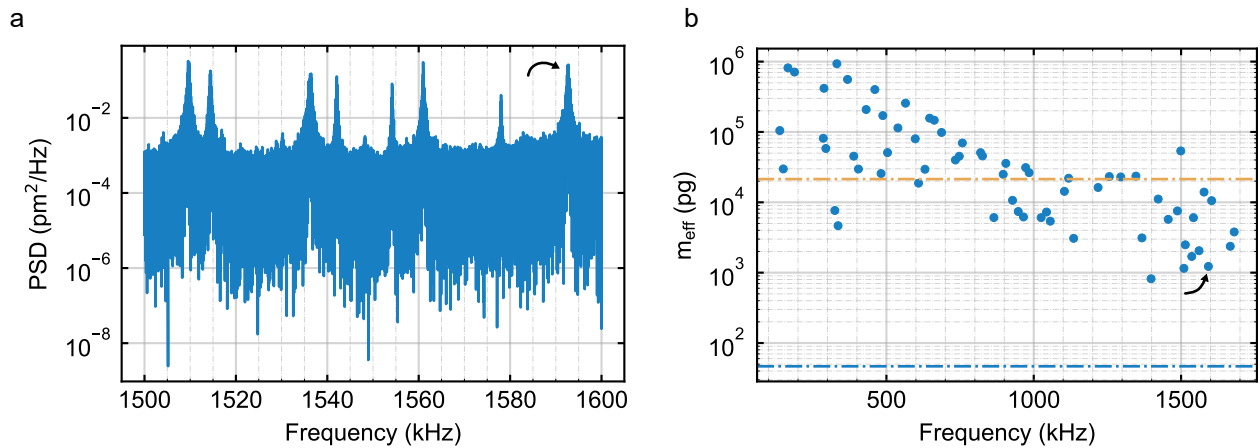

Figure S5: **Mechanical spectrum inside of the Fabry-Pérot microcavity**. (a) Displacement PSD for the modes between 1500 kHz and 1600 kHz, measured at the center of the hBN circular resonator. (b) Effective mass versus frequency for the mechanical modes measured at the center of the hBN drum. The orange and blue dashed lines depict the theoretical effective mass of the fundamental mode of an ideal rectangular SiN and circular hBN resonators, respectively, with the geometrical parameters of the experiment. The black arrow in both figures highlights the mode used for radiation pressure backaction experiments.

# S4 Influence of the SiN membrane on the optomechanical couplings

To rule out an effect of the SiN membrane on the static optomechanical couplings, we design our sample with an additional empty hole of the same diameter (40 µm) used for reference measurements. Figure S6a, top panel, shows the cavity detuning $\Delta$ as a function

of the sample position when the empty hole is centered with the cavity mode. There is no modulation of the detuning with periodicity $\lambda/2$, indicating the absence of dispersive coupling. The constant drift is due to thermal drifts of the cavity. The SiN frame slightly modulates the linewidth (Figure S6a, bottom panel), probably due to scattering of the mode at the edges, but remains very close to the empty cavity linewidth of 18.5 MHz, with some scattering around that value.

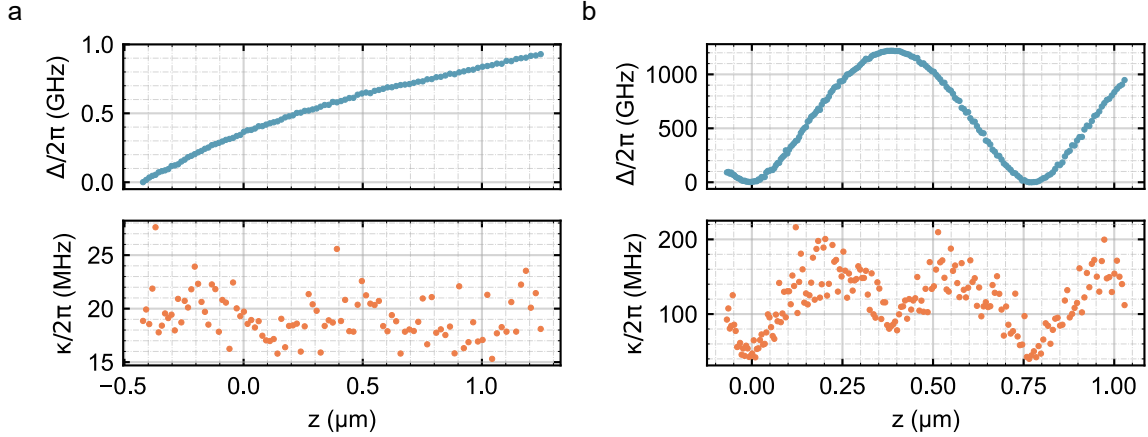

Figure S6: **Static spectroscopy.** (a) Cavity detuning $\Delta$ (top) and linewidth $\kappa$ (bottom) when the empty hole is placed at the cavity mode. (b) same as (a) but with the SiN membrane in place.

Figure S6b displays the detuning (top) and cavity linewidth (bottom) when the middle of the SiN membrane is centered with the cavity mode (notice the different scaling of the y-axis). The detuning, now periodic at $\lambda/2$, has a maximum value of $\Delta/2\pi = 1200\,\text{GHz}$. The cavity linewidth suffers a similar modulation as what is observed with the hBN flake. Neither the small absorption coefficient expected for hBN nor the larger one for SiN accounts for the large modulation observed on both materials. We believe it originates from the scattering losses to higher order modes. As discussed in the main text, this suggests that the magnitude of the modulation is not determined by the optical absorption of the material. We therefore attribute it to scattering losses into higher order optical modes, given a slight misalignment of the sample in the cavity (see the discussion in the main text for references).

# S5 Dispersive coupling: additional devices

Figure S7 shows the measured cavity detuning and dispersive optomechanical coupling for hBN drum-head resonators of different thicknesses. The thicknesses, displayed in the legend, were measured with an AFM. They are an upper limit to the real thickness as an additional $O_2$ plasma cleaning step etched the flakes, which was observed as a change of color contrast under the microscope. The thinnest flake ($< 16\,\text{nm}$) did not show a resolvable dispersive coupling due to its small reflectivity. The other two samples ($177\,\text{nm}$ and $24\,\text{nm}$) have a measurable dispersive coupling and show dynamical backaction. However, we did not characterize the degree of hybridization with the SiN membrane and consequently the data is not included in the main text.

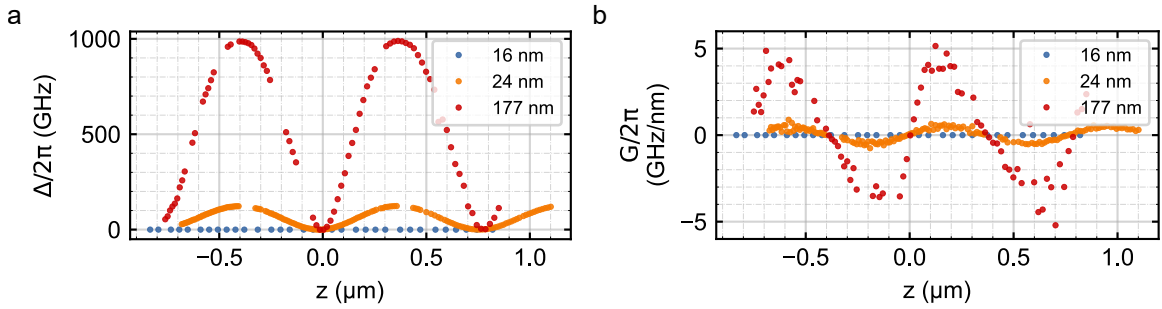

Figure S7: **Optomechanical coupling.** (a) Cavity detuning $\Delta$ and (b) dispersive optomechanical coupling $G$ for hBN resonators of different thicknesses (see legend).



\end{document}